\documentclass[journal,twoside,web]{ieeecolor}
\usepackage{generic}
\usepackage{cite}
\usepackage{amsmath,amssymb,amsfonts}
\usepackage{algorithmic}
\usepackage{graphicx}
\usepackage{textcomp}
\usepackage{float}
\usepackage{booktabs}
\usepackage{tikz}
\usepackage{cuted}
\usepackage{caption}
\usepackage[hidelinks]{hyperref}
\usepackage[switch]{lineno}
\def\BibTeX{{\rm B\kern-.05em{\sc i\kern-.025em b}\kern-.08em
    T\kern-.1667em\lower.7ex\hbox{E}\kern-.125emX}}
\begin{document}
\title{ Mechanisms of proton irradiation-induced defects on the electrical performance of 4H-SiC PIN detectors}
\author{Zaiyi. Li, Xiyuan. Zhang, Congcong. Wang, Haolan. Qu, Jiaxiang. Chen, Peilian. Liu, Suyu. Xiao, Xinbo. Zou, Hai. Lu and Xin. Shi 
\thanks{ This work is supported by the National Natural Science Foundation of China (Nos. 12205321,12375184,12305207 and 12405219), National Key Research and Development Program of China under Grant No. 2023YFA1605902 from the Ministry of Science and Technology, China Postdoctoral Science Foundation (2022M710085), Natural Science Foundation of Shandong Province Youth Fund (ZR2022QA098) under CERN RD50-2023-11  Collaboration framework. }
\thanks{Z. Li is with the Institute of High Energy Physics, Beijing 100049, China, also with the School of Physical Sciences, University of Chinese Academy of Sciences, Beijing 100049, China (e-mail: lizaiyi@ihep.ac.cn ). }
\thanks{X, Zhang, P. Liu, Congcong Wang, X. Shi is with the Institute of High Energy Physics, Beijing 100049, China, also with the State Key Laboratory of Particle Detection and Electronics, Beijing 100049, China (email: zhangxiyuan@ihep.ac.cn).}
\thanks{H, Qu, X. Zou is with School of Information Science and Technology (SIST), ShanghaiTech University, Shanghai 201210, China.}
\thanks{J. Chen was with School of Information Science and Technology (SIST), ShanghaiTech University, Shanghai 201210, China. He is now with Shenzhen Pinghu Laboratory, Shenzhen 518111, China. }
\thanks{H. Lu is with School of Electronic Science and Engineering, Nanjing University, Nanjing 210046, China.}
\thanks{S, Xiao is with Shandong Institute of Advanced Technology, Jinan, 1501, China.}
}

\maketitle

\begin{abstract}
Silicon Carbide (SiC) demonstrates significant potential for high-energy particle detection in complex radiation environments due to its exceptional radiation resistance, excellent environmental adaptability, and fast response time. Compared to silicon (Si) detectors, SiC detectors exhibit distinct radiation resistance characteristics depending on the type of radiation exposure. Here, the mechanism of the impact of 80MeV proton irradiation on the electrical performance of 4H-SiC PIN devices is reported.Deep Level Transient Spectroscopy (DLTS) and Time-Resolved Photoluminescence (TRPL) were utilized to analyze the defect characteristics and minority carrier lifetime of 4H-SiC detectors before and after irradiation, respectively. A Deep-Level Compensation Model (DLCM) was established using open source TCAD simulation tools RAdiation SEmiconductoR (RASER) to investigate mechanism responsible for the decrease in leakage current with rising radiation intensity, as well as the  constant-capacitance behavior exhibited  under proton irradiation up to $7.8\times10^{14} n_{eq}/cm^2$. The establishment of the physical model opens the door for the study of the influence mechanism of irradiation defects on SiC particle detectors.

\end{abstract}

\section{Introduction}
\IEEEPARstart{R}{adiation} detectors are of utmost significance in various fields, including nuclear medicine, nuclear power plant monitoring, particle physics, and related interdisciplinary areas. Compared with conventional gas detectors and scintillator detectors, semiconductor detectors developed in the 1960s and 1970s have the characteristics of high detection efficiency, simple structure, small size, fast response, high spatial and energy resolution \cite{I5}, \cite{I19}.  As a result, they have been more and more widely applied in the detection of $\alpha$, $\beta$, $\gamma$ and high-energy particles \cite{I1}, \cite{I6}. However, the properties of the intrinsic materials of semiconductor detectors determine that their detection performance will decline sharply with the increase of temperature and irradiation intensity. Therefore, they are not suitable for extreme environments such as reactors, high-energy physics research and deep space exploration \cite{I7}, \cite{I17}.  To overcome the limitations of the conventional detectors, the third generation wide band-gap semiconductor represented by SiC has the advantages of high breakdown voltage, fast carrier saturation drift velocity, high thermal conductivity, broader band gap and elevated atomic displacement threshold energy. In detail, the high breakdown voltage and fast carrier saturation drift velocity dedicate that SiC device are suitable for detecting high-event-rate signals and responding rapidly, which is more pronounced advantage in the field of high-energy particle detection \cite{I2}; the high thermal conductivity of SiC means that its operating temperature can be significantly increased, with a theoretical maximum operating temperature of up to 1240°C \cite{I16}; the broader band gap and elevated atomic displacement threshold energy indicate its strong resistance to radiation and prolonging the lifespan of detectors \cite{I15}.\par
With the commercialization of high-purity SiC materials and the increasing maturity of device processes, there have been numerous studies on the preparation, detection capabilities, radiation resistance, and performance degradation of SiC particle detectors \cite{I9}, \cite{I13}. Several studies have reported SiC detectors exposed to gamma ray, electron, neutron, and proton radiation \cite{I8}, \cite{I18}. As F.H.Rubby et al. and Akimasa Kinoshita et al. reported, $\gamma$ rays have virtually no impact on SiC detectors \cite{I3}, \cite{I4}. Electron irradiation including beta rays also has no noticeable effect on the device \cite{I10}. However, In terms of high-energy proton and neutron irradiation, high radiation doses can introduce a large number of deep-level defects, causing the radiation resistance properties of SiC to fall short of the expected levels \cite{I11}, \cite{I12}. It is worth noting that when focusing on radiation damage, studies usually only focus on the types and concentrations of defects produced; when studying the degradation of device performance, the main concern is the change of leakage current and charge collection efficiency. In addition, in the process of studying device performance and its degradation, it often relies on simulation or theoretical calculation to analyze the change of material property parameters, and lacks the actual measurement of related parameters. \par
In this work, the defect characteristics and minority lifetime before and after irradiation were measured, and the electrical properties of 4H-SiC PIN devices were studied by using the self-developed simulation software RASER \cite{I21}. The Deep Level Compensation Model (DLCM) for the 4H-SiC particle detector after high energy proton irradiation is proposed for the first time to describe the electrical properties. In summary, experiments combined with simulation analyzed and explained the degradation mechanism of the 4H-SiC particle detector from the perspective of defect generation and carrier lifetime degradation. This research contributes to the development of defect models, providing theoretical support for future device optimization and failure analysis.

\section{Samples and Experimental Setup}
\subsection{NJU-PIN diode and irradiation conditions}
To investigate the degradation mechanism of 4H-SiC diodes under irradiation, this study focus on the p+/n/n+ (PIN) diode  with two dimensions of $5 mm\ \times 5\ mm$ and $1.5mm\ \times1.5\ mm$, fabricated by Nanjing University. The active region of the PIN diode is a lightly doped N-type epitaxial layer with a thickness of approximately 100 $\mu m$ . A heavily doped P-type layer is epitaxially grown on top to serve as the anode, while the N-type doped conductive substrate acts as the cathode. The detailed structure can be seen from \cite{b}. The sample irradiation was conducted at the Associated Proton Beam Experiment Platform (APEP) beamline of the China Spallation Neutron Source (CSNS) in Dongguan, Guangzhou, China. The APEP facility delivers protons with an average energy of 80 MeV, and the fluence uncertainty is estimated to be below 10\%. The irradiation process was carried out in air at room temperature. Detailed information regarding the samples and irradiation conditions is presented in TABLE \ref{tab:samples}. Part of the test results of the $5\ mm\times5\ mm$ diodes were reported in \cite{a}. Detailed information regarding the samples and irradiation conditions is presented in TABLE \ref{tab:samples}. 
\begin{table}[H]
    \centering
    \caption{Information of PIN samples and irradiation fluences }
    \begin{tabular}{ccc}
    \toprule[1mm]
     label    & size ($mm^2$)& irradiation fluence ($n_{eq}/cm^2$)\\
     \midrule[1.5pt]
       NJU-PIN-1  & $1.5\times1.5$ & 0\\
       NJU-PIN-2  & $1.5\times1.5$ & $2\times10^{11}$\\
       NJU-PIN-3  & $1.5\times1.5$ & $3.5\times10^{12}$\\
       NJU-PIN-4  & $1.5\times1.5$ & $1\times10^{13}$\\
       NJU-PIN-5  & $1.5\times1.5$ & $1\times10^{14}$\\
       NJU-PIN-6  & $5\times5$ & 0\\
       NJU-PIN-7  & $5\times5$ & $3.9\times10^{13}$\\
       NJU-PIN-8  & $5\times5$ & $2.3\times10^{14}$\\
       NJU-PIN-9  & $5\times5$ & $7.8\times10^{14}$\\
    \bottomrule[1mm]
    \end{tabular}
    
    \label{tab:samples}
\end{table}

\subsection{Characterizations}

The current-voltage (I-V) characteristic was measured on a probe station at room temperature, using a Keithly 2470 source meter. The capacitance-voltage (C-V) characteristic was measured under same environment with Keysight E4980 LCR meter and a bias adapter, at frequency f=10kHz. Deep-level transient spectroscopy (DLTS) measurement was carried out utilizing the FT 1230 HERA DLTS system. Time-resolved photoluminescence (TRPL) measurement was carried out utilizing the FluoTime 300 PL spectrometer with a FluoMic PL microscope add-on.

\section{Measurement results}

\begin{figure}[]
    \centering
    \begin{minipage}[]{0.8\linewidth}
    \centerline{\includegraphics[width=\textwidth]{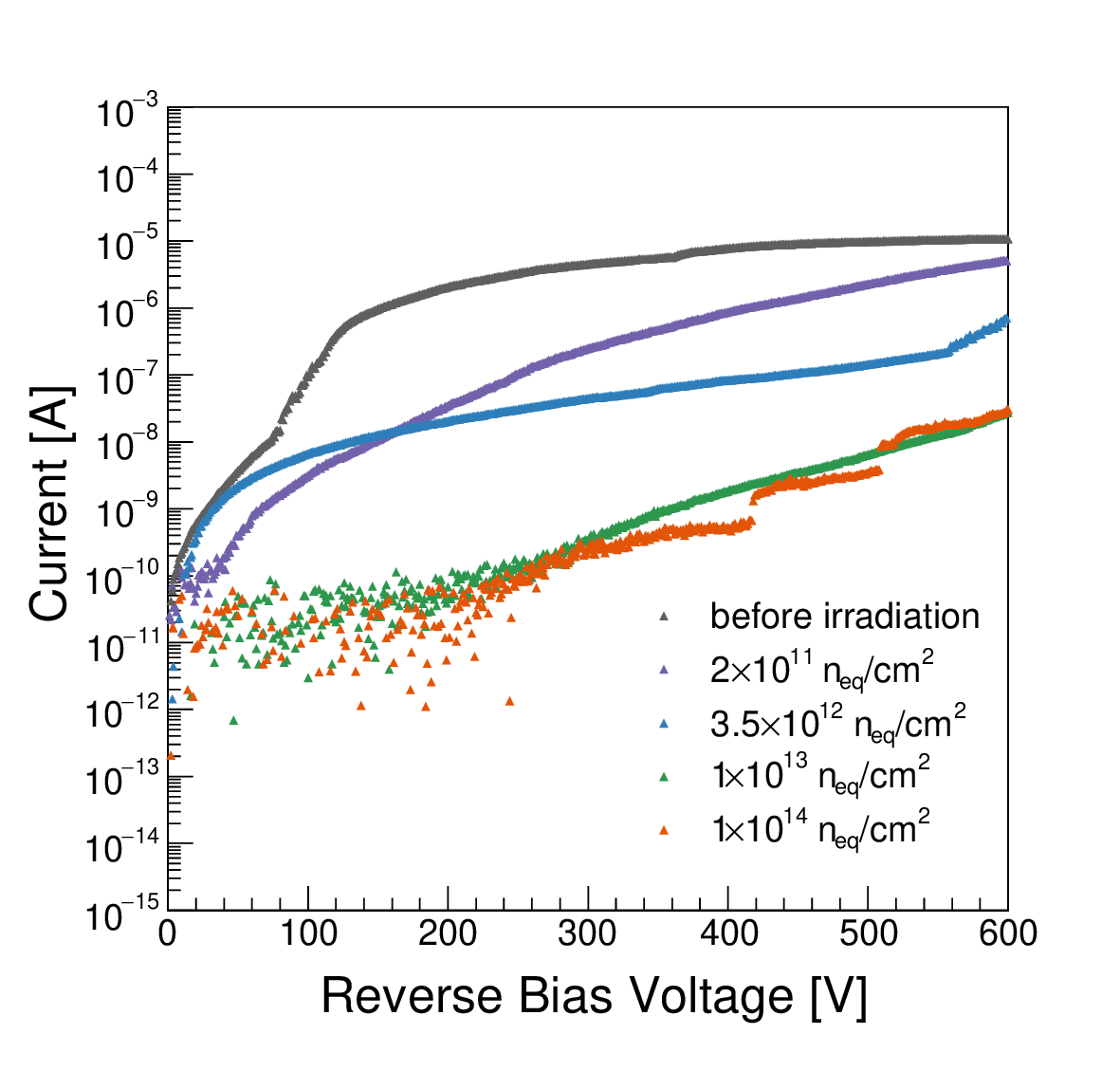}}
    \centerline{(a)}
    \end{minipage}
    \begin{minipage}[]{0.8\linewidth}
    \centerline{\includegraphics[width=\textwidth]{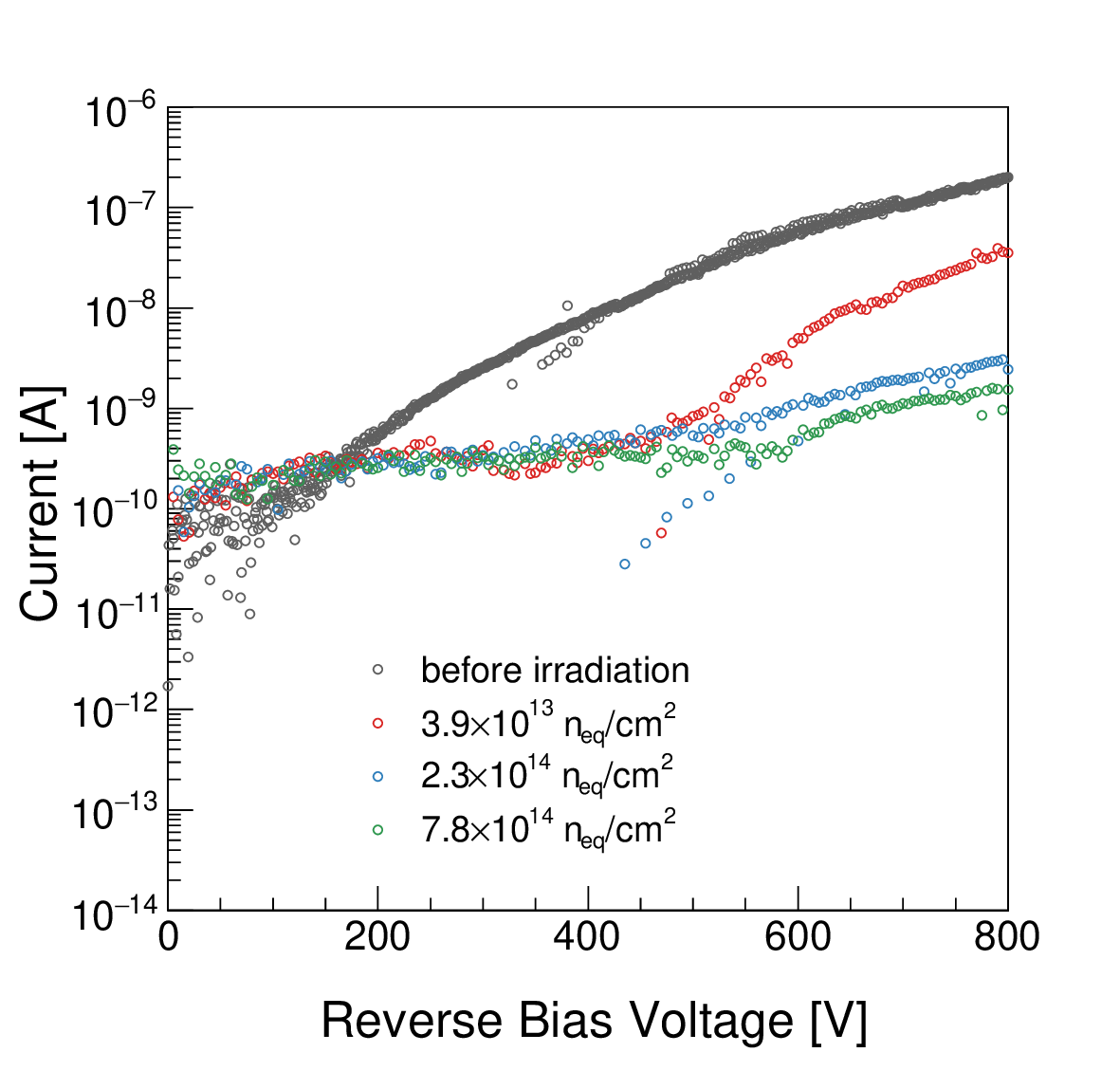}}
    \centerline{(b)}
    \end{minipage}
    \caption{Measured I-V characteristics of irradiated $1.5\ mm\times1.5\ mm$ (a) and $5\ mm\times5\ mm$ (b) diodes}
    \label{iv-data}
\end{figure}
\begin{figure}
    \centering
    \begin{minipage}[]{0.8\linewidth}
    \centerline{\includegraphics[width=\textwidth]{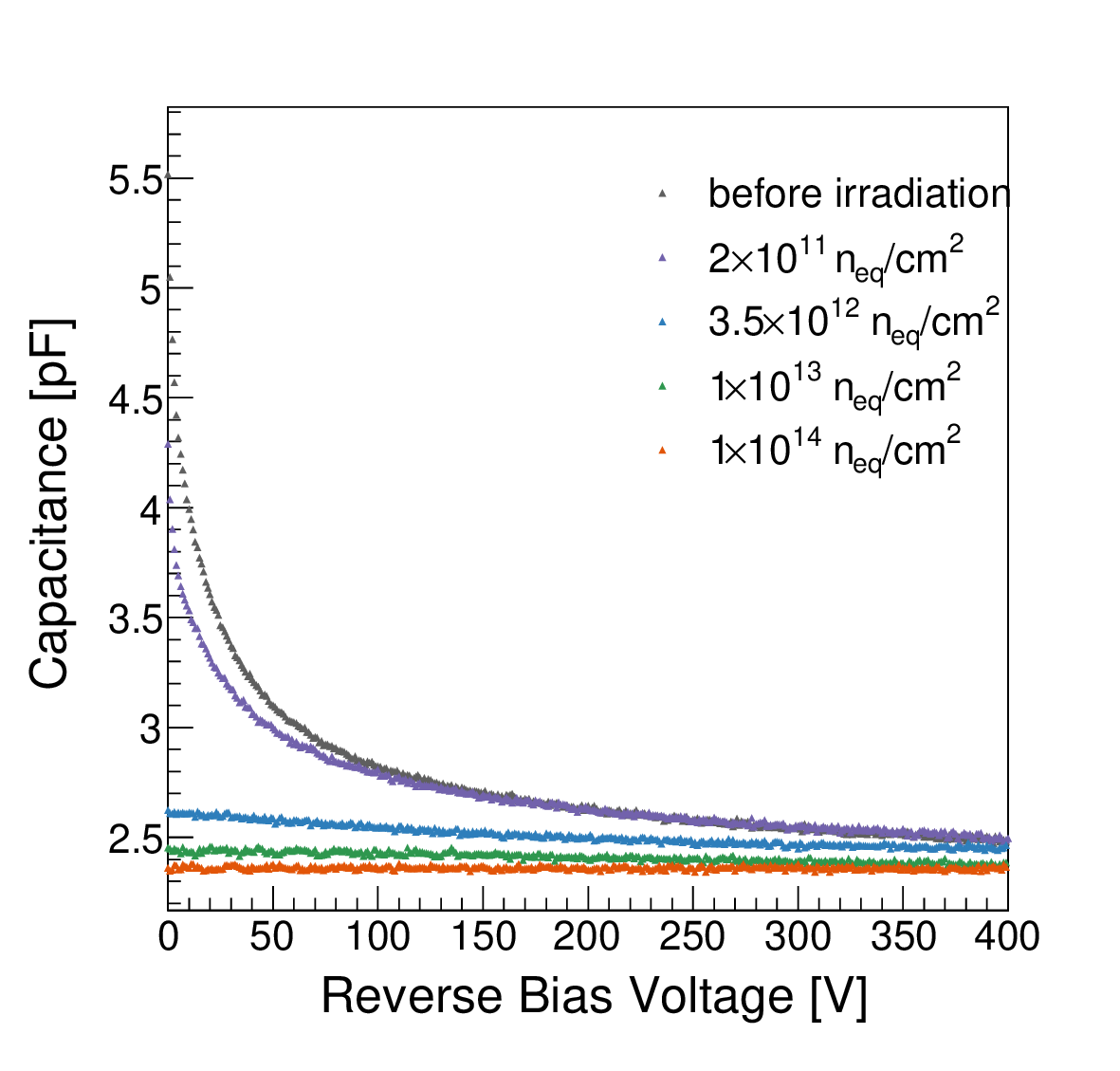}}
    \centerline{(a)}
    \end{minipage}
    \begin{minipage}[]{0.8\linewidth}
    \centerline{\includegraphics[width=\textwidth]{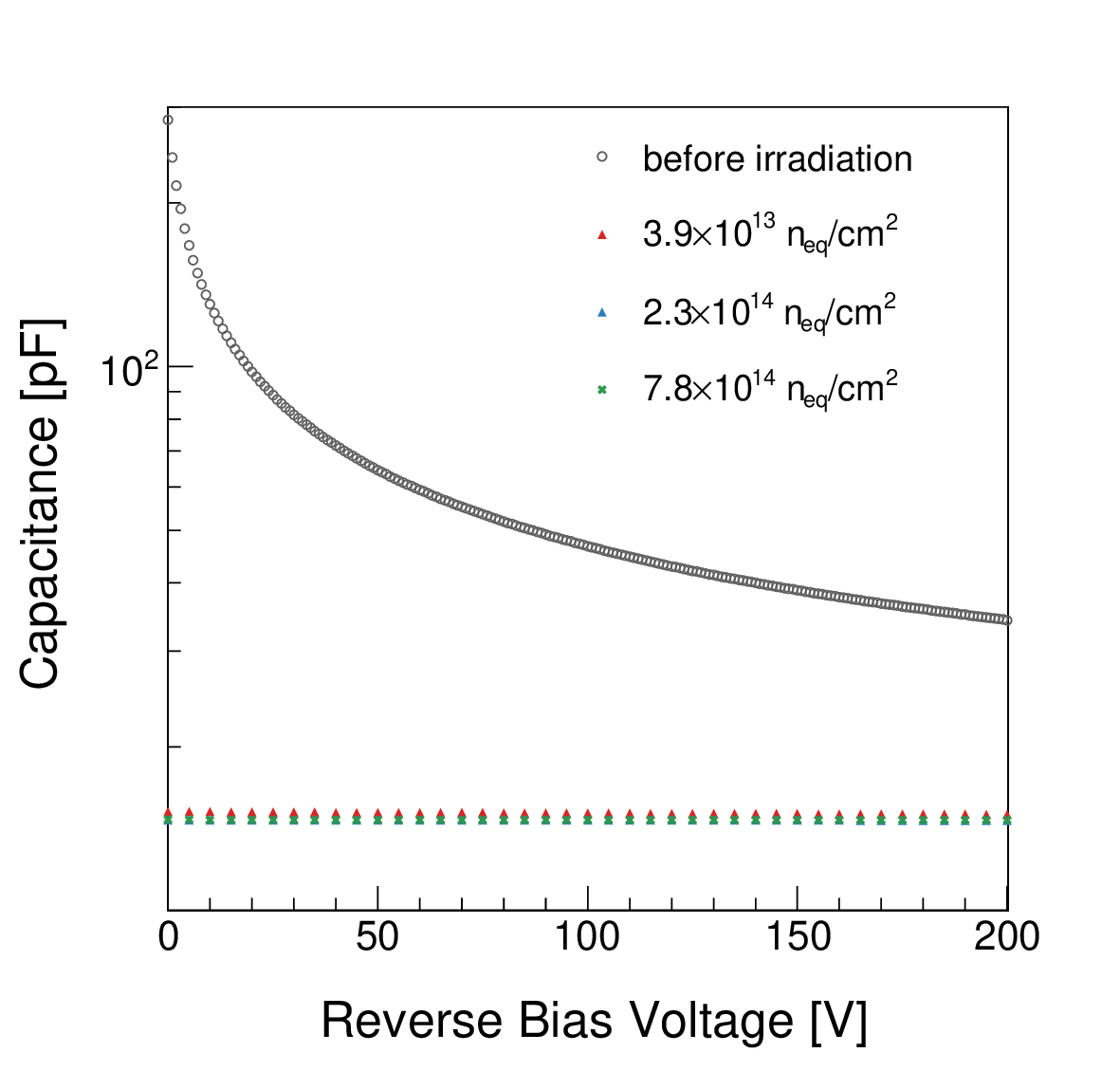}}
    \centerline{(b)}
    \end{minipage}    
    \caption{Measured C-V characteristics of irradiated $1.5\ mm\times1.5\ mm$ (a) and $5\ mm\times5\ mm$ (b) diodes}
    \label{cv-data}
\end{figure}

\subsection{I-V and C-V characteristics}
I-V characteristics reveal the voltage range in which device operates, determined by the equivalent noise level arising from leakage current prior to breakdown.
Fig. \ref{iv-data} presents the I-V characteristics of (a) $1.5\ mm\times1.5\ mm$ and (b) $5\ mm\times5\ mm$ 4H-SiC diodes after exposure to various proton irradiation fluences. The results demonstrate a significant leakage current reduction, with both device sizes showing a three-order-of-magnitude decrease after maximum irradiation. The change of leakage current after irradiation differs from the results observed for Si devices, where leakage currents are generally found to be approximately four to six orders of magnitude higher under room temperature conditions \cite{M1}. The observed jitter in the data points under low bias voltage can be attributed to the measurement resolution limit. The increased leakage current observed in the $1.5\ mm\times1.5\ mm$ devices compared to the $5\ mm\times5\ mm$ devices stems from the lack of rounded termination etching, a critical process for mitigating localized breakdown effects. Consequently, the $5\ mm\times5\ mm$ devices can be more accurately described as an ideal uniform PN junction and will be further investigated in the modeling section.\par
C-V characteristic reveals the effective doping concentration and illustrates how the depletion width varies in response to the applied voltage. 
Fig.\ref{cv-data} shows the C-V characteristics of $1.5\ mm\times1.5\ mm$ (a) and $5\ mm\times5\ mm$ (b) diodes measured under reverse bias. The non-irradiated diodes exhibit C-V characteristics typical of a PN junction. According to the $1/C^2-V$ curve derived from Fig. \ref{cv-data}(a), the $1.5\ mm\times1.5\ mm$ diode shows non-uniform doping, making it difficult to accurately determine doping concentration. The non-irradiated $5\ mm\times5\ mm$ diode shows full depletion at approximately 550 V, consistent with the epi-layer width of 100 $\mu m$. The effective doping concentration is determined to be $N_{eff}=5.2\times10^{13}\ cm^{-3}$. As the irradiation fluence increases, the C-V curve for the $1.5\ mm\times1.5\ mm$ diodes gradually becomes flat, while only the diode irradiated at $2\times10^{11}\ n_{eq}/cm^2$ remains diode-like. All the irradiated $5\ mm\times5\ mm$ diodes exhibit a flat C-V curve. The lower capacitance value at the same bias voltage indicates the diodes are more depleted after irradiation, indicating a reduction in doping concentration. The flat C-V curve suggests that the carrier concentration has diminished in the n-type region, causing it to behave like an intrinsic semiconductor. This phenomenon is attributed to the compensation caused by radiation-induced acceptor-like defects \cite{M4}.\par
 The capacitance of a fully depleted diode, which can be modeled as a parallel capacitor, is given as follow:
\begin{equation}
    C=\frac{\epsilon_sA}{W_D}
\end{equation}
Where A is the area of the planar device and $\epsilon_s$ is the dielectric constant of 4H-SiC, $W_D$ is depletion width. Using the measured constant capacitance value of 2.4 pF, the depletion width of the $1.5\ mm\times1.5\ mm$ device irradiated at $1\times10^{14}\ n_{eq}/cm^2$ is calculated to be 80 $\mu m$, where full compensation occurs. For the $5\ mm\times5\ mm$ diode irradiated at $7.8\times10^{14}\ n_{eq}/cm^2$, the constant capacitance of 14.7 pF observed from the flat C-V curve corresponds to a depletion width of 147 $\mu m$, which exceeds the width of the epi-layer. A detailed discussion of this phenomenon will be presented in \ref{section:cv-simulation}.\par
\subsection{Majority carrier traps}\label{section:DLTS}
The aforementioned I-V characterization demonstrates a reduction in leakage current under irradiation. Meanwhile, C-V measurements reveal that the devices gradually lose the diode-like characteristics with increasing irradiation fluence.  These changes in electrical properties may be related to the generation of deep-level defects. Taking irradiated silicon devices as an example, multiple deep-level defects have been introduced as an analytical approach \cite{D3}, \cite{D4}. Multiple deep levels that affect device electrical properties have been reported in irradiated 4H-SiC, including $EH_1$, $Z_{1/2}$, $EH_3$ and $EH_5$ \cite{D1}, \cite{D2}.\par
\begin{figure}[]
    \centering
    \begin{minipage}[]{1.0\linewidth}
    \centerline{\includegraphics[width=\textwidth]{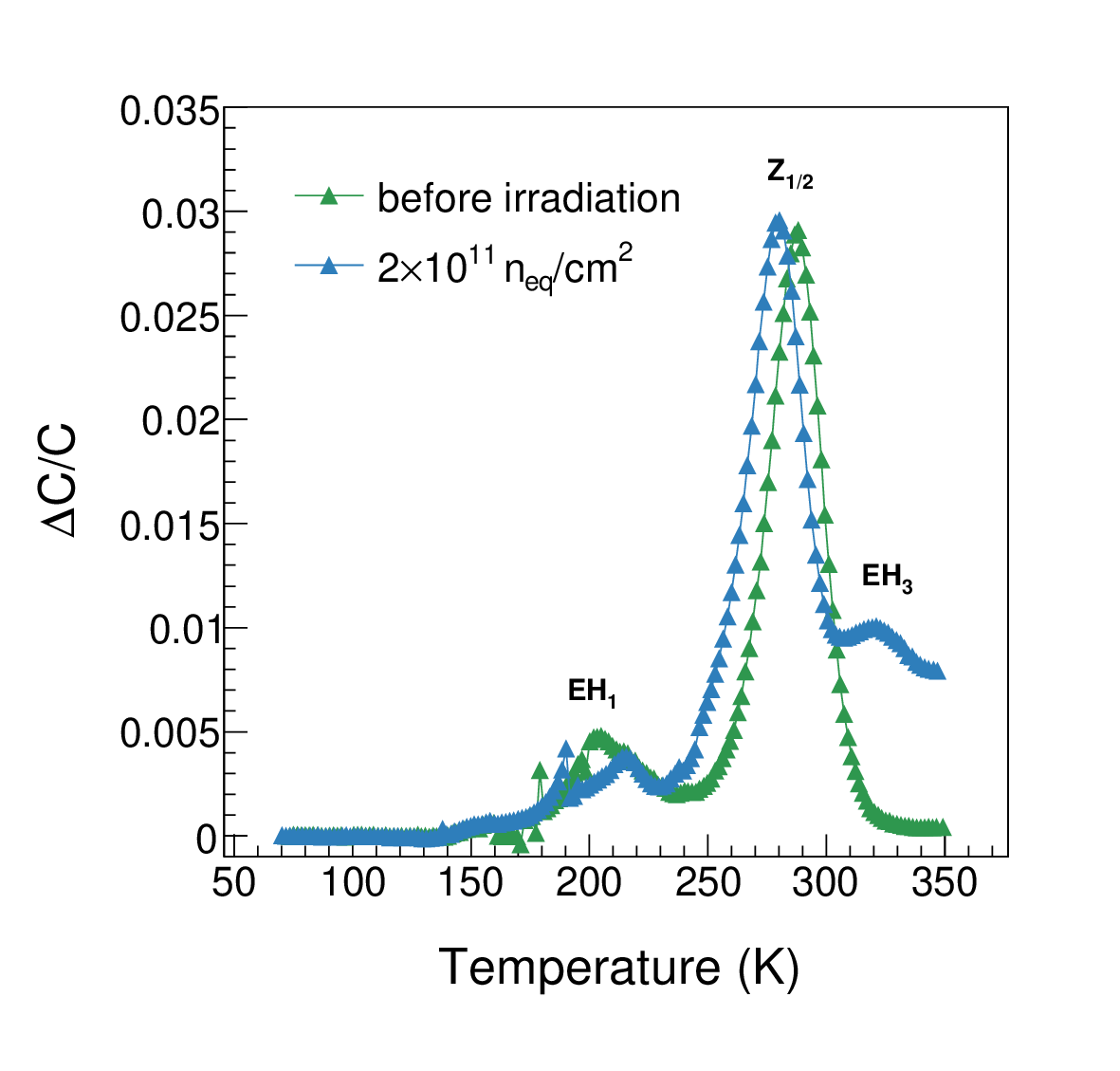}}
    \caption{DLTS signal observed before and after $2\times10^{11}n_{eq}/cm^2$ irradiation}
    \label{fig:dlts}
    \vspace{1em}
    \end{minipage}    
    \begin{minipage}[]{1.0\linewidth}   
    \centering   
    \captionof{table}{Electron Traps observed before and after $2\times10^{11}n_{eq}/cm^2$ irradiation}
    \setlength{\tabcolsep}{1mm}
    \begin{tabular}{ccccc}
    \toprule[1mm]
     label& $E_C-E_T(eV)$ & $N_{initial}(cm^{-3})$&$N_{irradiated}(cm^{-3})$&assignment\\
     \midrule[1.5pt]
     $EH_1$ & 0.44 &$1.40\times10^{11}$&$1.11\times10^{11}$&$V_{Si}$\\
     $Z_{1/2}$ & 0.63 &$8.52\times10^{11}$ & $8.66\times10^{11}$ &$V_C$\\
     $EH_3$ & 0.685 &-&$2.95\times10^{11}$& $V_{Si}$\\
     \bottomrule[1mm]
    \end{tabular} 
    \label{tab:dlts}
    \end{minipage}
\end{figure}

Fig. \ref{fig:dlts} shows the DLTS spectra of the non-irradiated and $2\times10^{11\ }n_{eq}/cm^2$ irradiated diodes. Since DLTS is not applicable to diodes whose capacitance does not change under pulse voltage, only the non-irradiated and $2\times10^{11}\ n_{eq}/cm^2$ irradiated diodes were tested. The bias voltage applied to the non-irradiated and irradiated samples is -8 V and -9 V, respectively, while a pulse voltage of 1 V is applied to both samples. In the pre-irradiated sample, two defect levels are identified between 70 K and 350 K: the $Z_{1/2}$ level (0.63 eV below $E_C$) and the $EH_1$ level (0.44 eV below $E_C$). Compared to the pre-irradiated sample, the post-irradiation trap levels exhibit slight energy shifts, suggesting defect modification under irradiation. Three trap levels emerge at 0.529 eV ($EH_1$), 0.607 eV ($Z_{1/2}$), and 0.685 eV ($EH_3$) below the conduction band edge ($E_C$). The jitter observed in the DLTS signal between 170 V and 190 V is  attributed to the probe lifting during temperature increase. Table. \ref{tab:dlts} summarizes the trap parameters for the majority carriers depicted in Fig. \ref{fig:dlts}.\par
The $EH_1$ defect is attributed to the silicon vacancy ($V_{Si}$) \cite{D5}. After irradiation, the observed concentration decreases slightly. The increased bias voltage in irradiated samples indicates deeper probing region from the surface, suggesting higher defect density in near-surface regions. Therefore, the $EH_1$ defect is more likely associated with the fabrication process, such as ion implantation. The $Z_{1/2}$ level is attributed to the double acceptor (0/2-) transition of the carbon vacancy ($V_C$), which exhibited significant increase upon irradiation \cite{D6}. However, the signal peak of the $Z_{1/2}$ defect only shows a slight increase compared to the non-irradiated sample. Due to the experimental uncertainties associated with the DLTS test, the increase of the defect $Z_{1/2}$ after irradiation, as noted by other studies, cannot be definitively confirmed. It is possible that the concentration of the $Z_{1/2}$ defects generated by irradiation is comparable to or lower than the concentration prior to irradiation. The $EH_3$ defect is attributed to the acceptor state of silicon vacancy($V_{Si}$) \cite{D5}. In contrast to the $EH_1$ and $Z_{1/2}$ defects, which does not exhibit a significant increase following irradiation, the $EH_3$ defect is uniquely observed in the irradiated sample. This observation identifies the $EH_3$ defect as a key factor in the degradation of electrical properties, particularly concerning the previously mentioned compensation effect on the donors. Conversely, the $EH_1$ and $Z_{1/2}$ defects do not appear to significantly contribute to this degradation.\par

\subsection{Impact of radiation damage on minority carrier lifetime}
\begin{figure}[]
    \centering
    \begin{minipage}[]{1.0\linewidth}
    \centerline{\includegraphics[width=\textwidth]{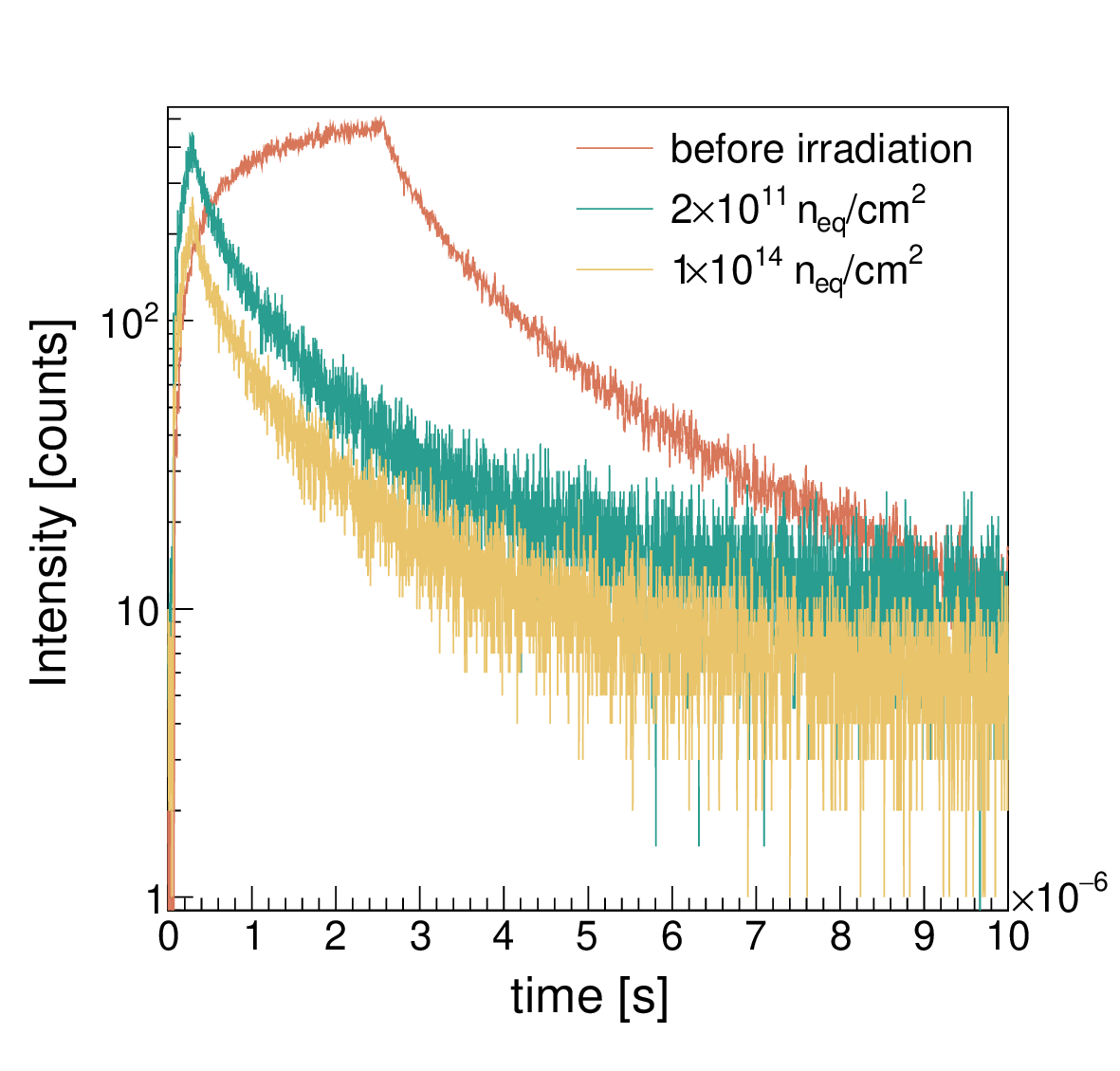}}
    \caption{TRPL transient signal intensity before irradiation and after $2\times10^{11} n_{eq}/cm^2$,$1\times10^{14} n_{eq}/cm^2$ irradiation. The data in the figure have been scaled to prevent overlap.}
    \label{fig:data-trpl}
    \vspace{1em}
    \end{minipage}
    \begin{minipage}[]{1.0\linewidth}   
    \centering   
    \captionof{table}{Minority carrier lifetime achieved from TRPL transients}
    \setlength{\tabcolsep}{1mm}
    \begin{tabular}{cccc}
    \toprule[1mm]
     $\Phi(n_{eq}/cm^2)$    & 0 &$2\times10^{11}$& $1\times10^{14}$ \\
     \midrule[1.5pt]
      $\tau$(ns)   & 484 &398&376\\
    \bottomrule[1mm]
    \end{tabular} 
    \label{tab:tau}
    \end{minipage}
\end{figure}

The irradiation process outlined above results in the generation of defect $EH_3$. In principle, a higher density of minority lifetime-killing defects is expected to contribute to a reduced minority carrier lifetime \cite{TR1}. To clarify this issue, we conducted TRPL measurements, which offer direct insight into the carrier lifetime within the epi-layer. The typical TRPL transients obtained from different samples are exemplified in Fig. \ref{fig:data-trpl}. Multi-component time parameters are achieved by fitting exponential function $\eqref{eq-trpl}$ to data and the minority carrier lifetime in the epi-layer is summarized in Table. \ref{tab:tau}.
\begin{equation}
    Dec(t)=\sum_i^{n_{Exp}}A_ie^{-\frac{t}{\tau_i}}+Bkg\label{eq-trpl}
\end{equation}
\par The irradiation at fluences of $2\times10^{11}\ n_{eq}/cm^2$ and $1\times10^{14}\ n_{eq}/cm^2$ resulted in minority carrier lifetime reductions of 18\% and 22$\%$, respectively, compared to the pre-irradiation sample. The experimental verification confirms the appropriateness of the 500 ns hole lifetime parameter implemented in RASER prior to irradiation.
\section{Modeling of proton irradiated diodes}
As mentioned earlier, the deep-level defects in 4H-SiC detectors undergo changes after proton irradiation. Deep-level defects typically impact device performance in the following two aspects: 
compensation effect and the impact of recombination centers on minority carrier lifetime, ultimately leading to the degradation of device performance. This may explain the significant variations in electrical performance observed in the previous discussion. To further investigate the impact mechanism of deep-level defects on device degradation, TCAD simulations were conducted for $5\ mm\times5\ mm$ devices based on RASER. All simulations have been conducted out at room temperature (300K).
\subsection{Compensation effect}
Electrically active deep-level defects contain highly localized charges, which can behave like acceptors or donors, thereby affecting the nearby carrier concentration, resulting in the change in electric field distribution.
The compensation effect induced by deep-level defects lowers the effective doping concentration through the capture of charge carriers, driving the Fermi level toward the mid-gap. When the deep-level defect concentration reaches a critical threshold, the Fermi-level pinning effect stabilizes the Fermi level at the mid-gap \cite{T1}. This mechanism leads to a significantly reduced effective doping concentration in the depletion region, resulting in a voltage-independent depletion width and intrinsic semiconductor-like behavior, consistent with the observed C-V characteristics. In our n-type 4H-SiC device, acceptor-like defects contribute to the compensation effect. Our DLTS test confirms the acceptor nature of $EH_3$ generated following irradiation. \par
The compensation effect describes the mutual neutralization of carrier concentrations by coexisting donor impurities and acceptor-like defects, where donors supply electrons and acceptors capture them, reducing the effective doping concentration. Therefore, the concentration of generated defects equals to the change in effective doping concentration and is proportional to the irradiation fluence, with a coefficient g, expressed as follow:
\begin{equation}
  \Delta N_{eff}=\Delta D=g \Phi_{n_{eq}} \label{eq-g-def} 
\end{equation}
Where $\Phi_{n_{eq}}$ is the neutron-equivalent irradiation fluence, D is the defect concentration, $N_{eff}$ is effective doping concentration, g is the linear introduction rate of defects with radiation fluence. 
\subsection{Application to C-V simulation}\label{section:cv-simulation}
The C-V characteristics are obtained by solving the Poisson equation in appendix. According to \ref{section:DLTS}, only the defect $EH_3$ shows a significant increase after irradiation. Therefore, DLCM considers solely $EH_3$ defects, with doping concentration changes $\eqref{eq-g-def}$ serving as charge density in the irradiated Poisson equation. The $EH_3$ defect introduction rate is determined as $g\approx1.48 cm^{-1}$ for the n-type region from Table. \ref{tab:dlts}. Notably, in 4H-SiC, radiation-induced defects exhibit dual compensation of both donor and acceptor species. For the p+/n/n+ diode structure under study, the deep-level introduction rate in the p-type region is assigned g=1.0 $cm^{-1}$, demonstrating negligible impact.\par
Accurate structural parameter definition is critical for reliable device behavior analysis when solving the Poisson equation. In this simulation, we focus on a one-dimensional p+/n/n+ structure, where the n region has a doping concentration of approximately $N_D\approx5.2\times10^{13}\ cm^{-3}$ . 
The effective doping concentration is obtained from the C-V data before irradiation and is consistent with the value reported in \cite{b}. Before irradiation, the width of the epi-layer is 100 $\mu m$. After irradiation, the structure is adjusted slightly based on the test results. When a width of 100 $\mu m$ is applied for the epi-layer, the simulated full-depleted capacitance is found to be 21.6 pF, which deviates from the measured value of 14.7 pF. The width of the depletion region, corresponding to the flat C-V curve observed after irradiation, is calculated from the data to be 147 $\mu m$. To obtain an accurate electric field, this equivalent additional structure must not be overlooked following irradiation. The substrate layer appears to have experienced a compensation effect that reduced its doping concentration and extended the depletion region, mirroring  to the compensation effect observed in heavily doped GaN regions following high irradiation fluence \cite{d}. Accordingly, the following equivalent modifications were implemented in the structural model simulation, a region approximately 47 $\mu m$ thick with a doping concentration of $N_D\approx3.0\times10^{14}\ cm^{-3}$ is added to the end of the epi-layer to align with the test result.\par
\begin{figure}
    \centering
    \includegraphics[width=1.0\linewidth]{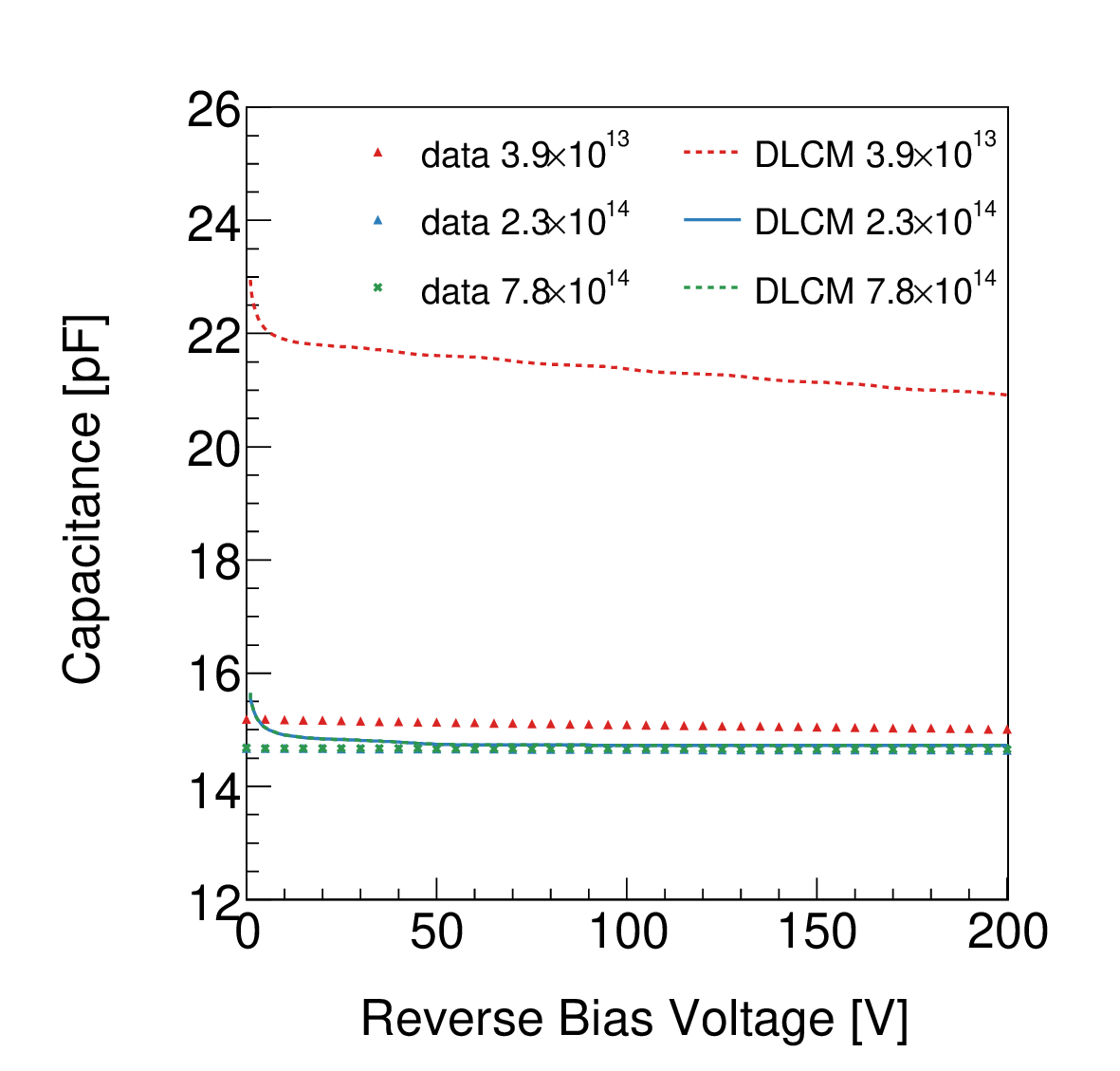}
    \caption{Measured and simulated C-V by DLCM for diodes irradiated from 0 to $7.8\times10^{14} n_{eq}/cm^2$ }
    \label{fig:cv-sim}
\end{figure}

Fig. \ref{fig:cv-sim} shows the simulation result of capacitance-voltage characteristics using DLCM. Before irradiation, the simulated C-V perfectly matches the data. For higher irradiation fluences the simulation matches the measurements quite well over the fluence range whereas for lower fluence a small deviation is observed. This is due to the uncertainty of the epi-layer/substrate structure and irradiation fluence. The agreement between the data and the simulation suggests that the assumptions in the DLCM is appropriate for the irradiated devices.\par
\begin{figure}[]
    \centering
    \includegraphics[width=1.0\linewidth]{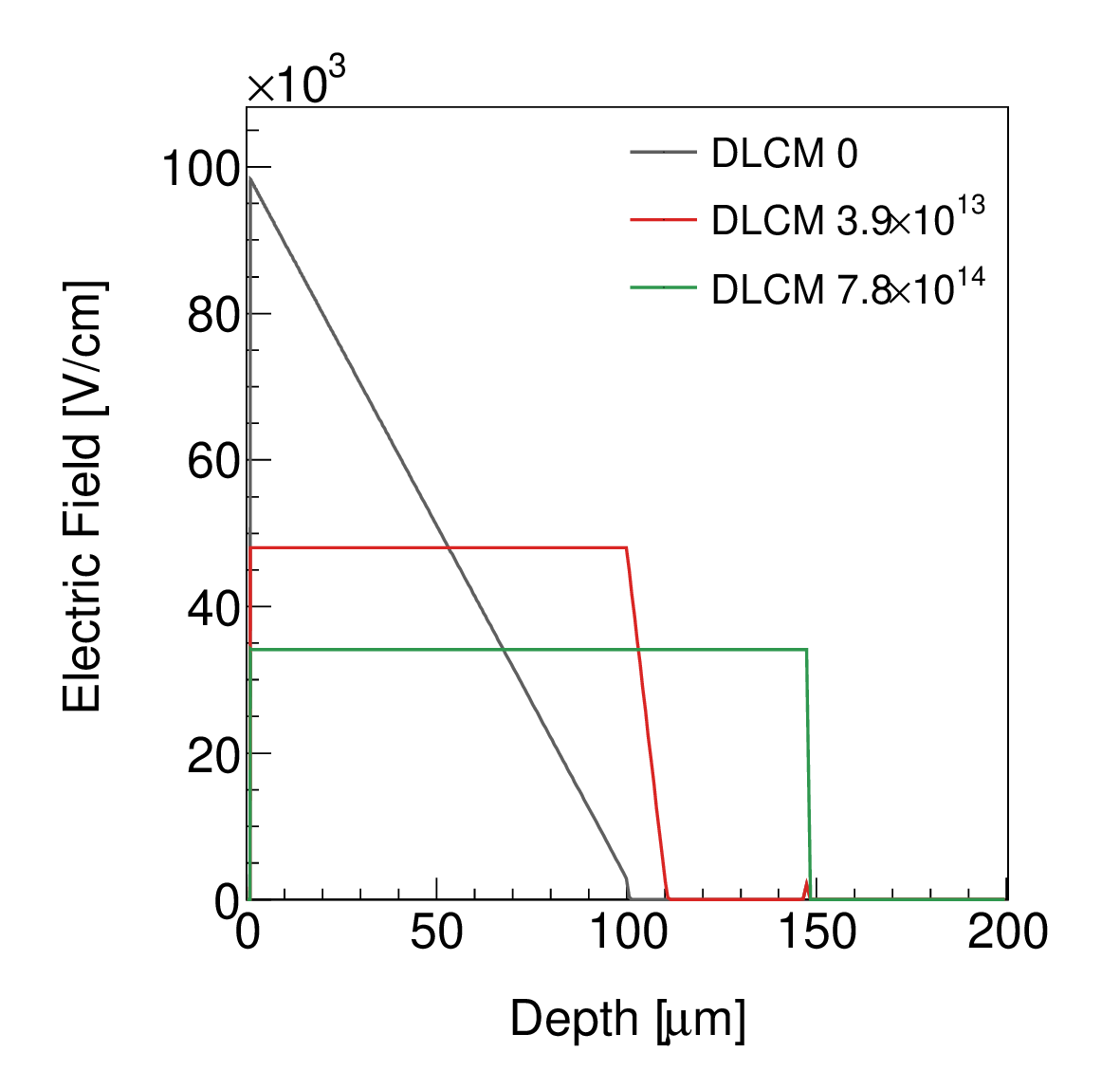}
    \caption{electric field inside irradiated diodes simulated by DLCM for diodes irradiated from 0 to $7.8\times10^{14} n_{eq}/cm^2$}
    \label{fig:sim-field}
\end{figure}

Solving Poisson equation also yields the electric field inside irradiated diodes. As Fig. \ref{fig:sim-field} shows, after irradiation, the distribution of the electric field within the device has changed due to the doping compensation effect. The maximum electric field decreases as the overall electric field gradually becomes uniformly distributed. The electric field can be further utilized to simulate the charge collection efficiency of irradiated devices, and the degradation of CCE will be assessed in future experiments. Although the occurrence of the double peak effect in silicon devices after high fluence irradiation is well recognized \cite{S6}, it is not included in this model due to a lack of experimental observations. Once this effect has been observed, it can be easily simulated in RASER, as the carrier trapping mechanism is already integrated into the software.
\subsection{Recombination process}
The reverse leakage current characteristics of both 4H-SiC and Si diodes can be understood within the framework of Shockley-Read-Hall (SRH) recombination theory \cite{T3}. In reverse bias, traps alternatively emit electrons and holes, forming a generation current proportional to the depletion width. In uniformly doped PN junctions, the depletion width $W_D$ is proportional to square root of the applied reverse bias voltage V, expressed as $W_D\propto V^{1/2}$. Therefore, the leakage current varies as $V^{1/2}$. Here, for simplicity, we assume that the generation current is generated over the whole depletion width. Thus the leakage current density according to SRH recombination is \cite{M2}:
\begin{equation}
    J_{SRH}\approx\frac{qn_iW_D}{\tau}\label{eq-J-diode}
\end{equation}
Where $W_D$ is the width of depletion region, $\tau$ is the lifetime of minority carriers, and $n_i$ is the intrinsic carrier concentration. The minority carrier recombination lifetime, $\tau$, represents the time constant for the recombination of non-equilibrium electron-hole pairs, or excess minority carriers. It also indicates the duration required for electron-hole pairs to generate and achieve equilibrium.\par
Equation $\eqref{eq-J-diode}$ indicates that the leakage current is proportional to the depletion width and will saturate once full depletion is achieved. The $5\ mm\times5\ mm$ diode reaches full depletion at approximately 550 V; 
 however, the leakage current in Fig. \ref{iv-data}(b) continues to increase significantly. 
 This phenomenon challenges the validity of conventional SRH theory in high-field regimes, indicating that the recombination model employed for our device requires optimization under high electric fields. In high electric field regions, impact ionization and tunneling effects may arise, with their relative contributions governed by the material's intrinsic properties.
When the electric field in 4H-SiC diodes exceeds $\sim$40 kV/cm, tunneling effects cause the leakage current characteristics to deviate from the ideal $V^{1/2}$ relationship.
The tunneling process of charge carriers can be equivalently treated as a generation process, namely the field-enhanced SRH recombination. The field-enhanced SRH recombination theory is expressed by \cite{M3}:
\begin{equation}
\begin{split}
    &\resizebox{1.\hsize}{!}{$R=\frac{np-n_i^2}{\tau_p(F)[n+n_iexp((E_t-E_i)/kT)]+\tau_n(F)[p+n_iexp((E_i-E_t)/kT)]}$}\\
    &\quad\approx\frac{n_i}{\tau(F)}\label{srh-definition1}
\end{split}
\end{equation}
Where $\tau_n$ and $\tau_p$ are the carrier lifetimes of electrons and holes, F is field strength, n and p are the electron and hole densities, $n_i$ is the intrinsic carrier concentration, $E_t$ and $E_i$ are the trap energy and the intrinsic Fermi level energy, respectively. The approximation is made by neglecting the carrier density in the depletion region. The only difference between $\eqref{srh-definition1}$ and conventional SRH theory is the field dependence of the lifetimes, expressed by $\eqref{eq-tau-F}$.\par
Considering the tunneling process, the generated current density considering the electric field dependence \cite{T2} can be expressed as
\begin{equation}
    J=(1+g(F))J_{SRH}
\end{equation}
Where g(F) accounts for the tunneling process. This is equivalent to applying $\eqref{eq-J-diode}$ while using a modified SRH lifetime 
\begin{equation}
    \tau(F)=\frac{\tau_0}{1+g(F)}\label{eq-tau-F}
\end{equation} expressed by means of field-enhancement factor g(F). Field-enhancement factors are implemented in RASER based on the tunneling via traps \cite{T2}.\par
The carrier lifetime in $\eqref{srh-definition1}$ can be approximately written as:
\begin{equation}
    \tau_{\alpha}\approx\frac{1}{\sigma_{\alpha}N_t}, \alpha=n,p
\end{equation}
Where $N_t$ is the trap concentration and $\sigma_{\alpha}$ is the capture cross-section of traps. Considering only the trap with the largest capture cross-section, the carrier lifetime decreases with increasing trap concentration. Assuming that the generated trap concentration is proportional to irradiation fluence, the carrier lifetime follows the relationship $\frac{1}{\tau}\propto\Phi_{n_{eq}}$. This relation has been validated in silicon by \cite{S4}. Based on our TRPL measurements of the minority carrier lifetime, we observe that $\tau_p$ decreases following irradiation due to an increase in traps; however, $\frac{1}{\tau}$ is not totally proportional to the irradiation fluence. The carrier lifetime, when compared to the fresh sample, decreased by 18\% after an irradiation fluence of $2\times10^{11}\ n_{eq}/cm^2$ and by only 22\% following an irradiation fluence of $1\times10^{14}\ n_{eq}/cm^2$, representing an increase in fluence by three orders of magnitude. This observation indicates a deviation from the SRH theory, which will be demonstrated in our simulation.
\subsection{Application to I-V simulation}\label{sec-iv-sim}
The field-enhanced SRH recombination $\eqref{srh-definition1}$ is applied as the generation-recombination term of the current continuity equations, with constant carrier mobilities ($\mu_n, \mu_p$). The field-enhance factor takes the form of the Hurkx model \cite{T2}.
The only tuning parameter $m^*=0.25\ m_e$, which equals to the value used in silicon. This parameter is tuned by the leakage current data before irradiation and is fixed after irradiation.
\begin{figure}[]
    \centering
    \begin{tikzpicture}
    \draw [ultra thick] (0,0) -- (3,3);
    \draw [ultra thick] (1,-2) -- (4,1);
    \draw [thick] (1.0,0.5) -- (1.4,0.5);
    \draw [thick] (2.5,0.5) -- (2.9,0.5);
    \draw [->] (2.7,-0.25) -- (2.7,0.47);
    \draw [->] (1.2,0.5) -- (1.2,1.1);
    \draw [->] (2.4,0.5) -- (1.5,0.5);
    \draw [->] (0.95,0.5) -- (0.55,0.5);
    \draw [->] (3.4,0.5) -- (2.95,0.5);
    \draw [->,thick] (0,-0.6) -- (1.8,-0.6);
    \node at (1.2,0.3) [ scale=0.75] {$E_1$};
    \node at (2.7,0.7) [ scale=0.75] {$E_2$};
    \node at (1.4,0.8) [ scale=0.75] {1};
    \node at (0.75,0.3) [scale=0.75] {2};
    \node at (2.0,0.65) [ scale=0.75] {3};
    \node at (2.55,0) [  scale=0.75] {4};    
    \node at (3.2,0.7) [scale=0.75] {5};
    \node at (0.9,-0.8) [scale=0.8] {Electric Field};
    \node at (3.3,3) {$E_C$};
    \node at (4.3,1) {$E_V$};
    \filldraw [color=gray] (1.2,1.31) circle (2.5pt);
    \filldraw [color=gray] (0.43,0.57) circle (2.5pt);
    \filldraw [color=gray,fill=white,thick] (2.7,-0.41) circle (2.5pt);
    \filldraw [color=gray,fill=white,thick] (3.58,0.42) circle (2.5pt);
    \end{tikzpicture}
    \caption{Electron generation process with a consideration of the energy band in the reverse-biased depletion region, path 1,4 is conventional SRH recombination, path 2,3,5 is tunneling under electric field}
    \label{fig:recombination-illustration}
\end{figure}
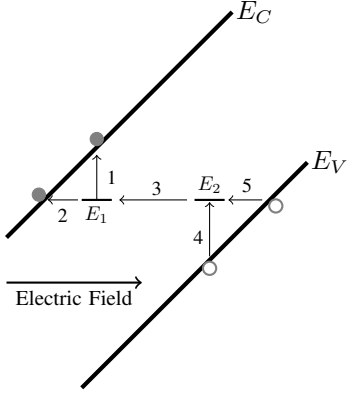
The TRPL results demonstrate the relationship between minority carrier lifetime degradation and irradiation fluence. The prediction of carrier lifetime under any irradiation fluence is done by fitting data in Table. \ref{tab:tau} to a linear function: $\frac{1}{\tau}=a\cdot ln\Phi_{neq}+b$. This is mathematically equivalent to multiplying the total SRH recombination rate by a factor $\alpha=\frac{a\cdot ln(\Phi_{neq}+1)+b}{b}$, as a result of carrier lifetime decrease, where $a=2.4\times10^4$ and $b=1.9\times10^6$. Therefore, when simulating another device with a significantly different lifetime $\tau$ compared to our device, the non-irradiated $\tau$ can be adjusted accordingly. \par
The intrinsic carrier concentration can be calculated by: 
\begin{equation}
    n_i=\sqrt{N_CN_V}exp(-\frac{E_g}{2kT})\label{ni-definition}
\end{equation}
Where $E_g$ is the bandgap width,  $N_C$, $N_V$ are effective state densities of conduction band and valence band. However, inserting the intrinsic carrier density calculated from $\eqref{ni-definition}$ and the measured $\tau\approx500$ ns into $\eqref{srh-definition1}$ for simulation, the measured current density is much larger than expected in $\eqref{eq-J-diode}$. Therefore, we preserve the formula while recalibrating the value of $n_i$ as effective intrinsic carrier density ($n_i^{eff}$).  According to the I-V data prior to irradiation, $n_i^{eff}$ is adjusted to be equal to $3\times10^5 cm^{-3}$.
Same method was applied in \cite{c}, in which pointing out a new generation mechanism in 6H-SiC, depending on temperature and doping concentration. According to our work, this generation mechanism also works for 4H-SiC. Experimental observation of the generation mechanism in 4H-SiC and 6H-SiC devices can be found in \cite{S3}.
A possible explanation of the generation mechanism is that in wide-bandgap semiconductors, the main recombination approach involves two defect levels, and the transition of the electron between the levels can be assisted by electric field, as illustrated in Fig. \ref{fig:recombination-illustration}. Detailed theoretical discussion has been made, with application to silicon carbide \cite{S1}. Another theoretical explanation is the metastable-structure-relaxation between different charged defect states, which has been demonstrated in silicon dioxide \cite{S2}, whose applicability to SiC requires verification through additional computational studies. Both references support the deviation from SRH theory.  
After heavy irradiation, the $n_i^{eff}$ is adjusted by multiplying a factor of 0.55 to match data. This can be explained by the fact that the carrier generation mechanism depends on doping concentration, since all the simulated samples behave like nearly intrinsic after irradiation.\par
Finally, the generation-recombination term can be written as:
\begin{equation}
    G-R=\alpha\cdot\frac{n^{eff}_i}{\tau(F)}\label{eq:G-R def}
\end{equation}
Where $\alpha$ is the product of modification coefficients, and $\tau$ is in the form of $\eqref{eq-tau-F}$.
\begin{figure}
    \centering
    \includegraphics[width=1.0\linewidth]{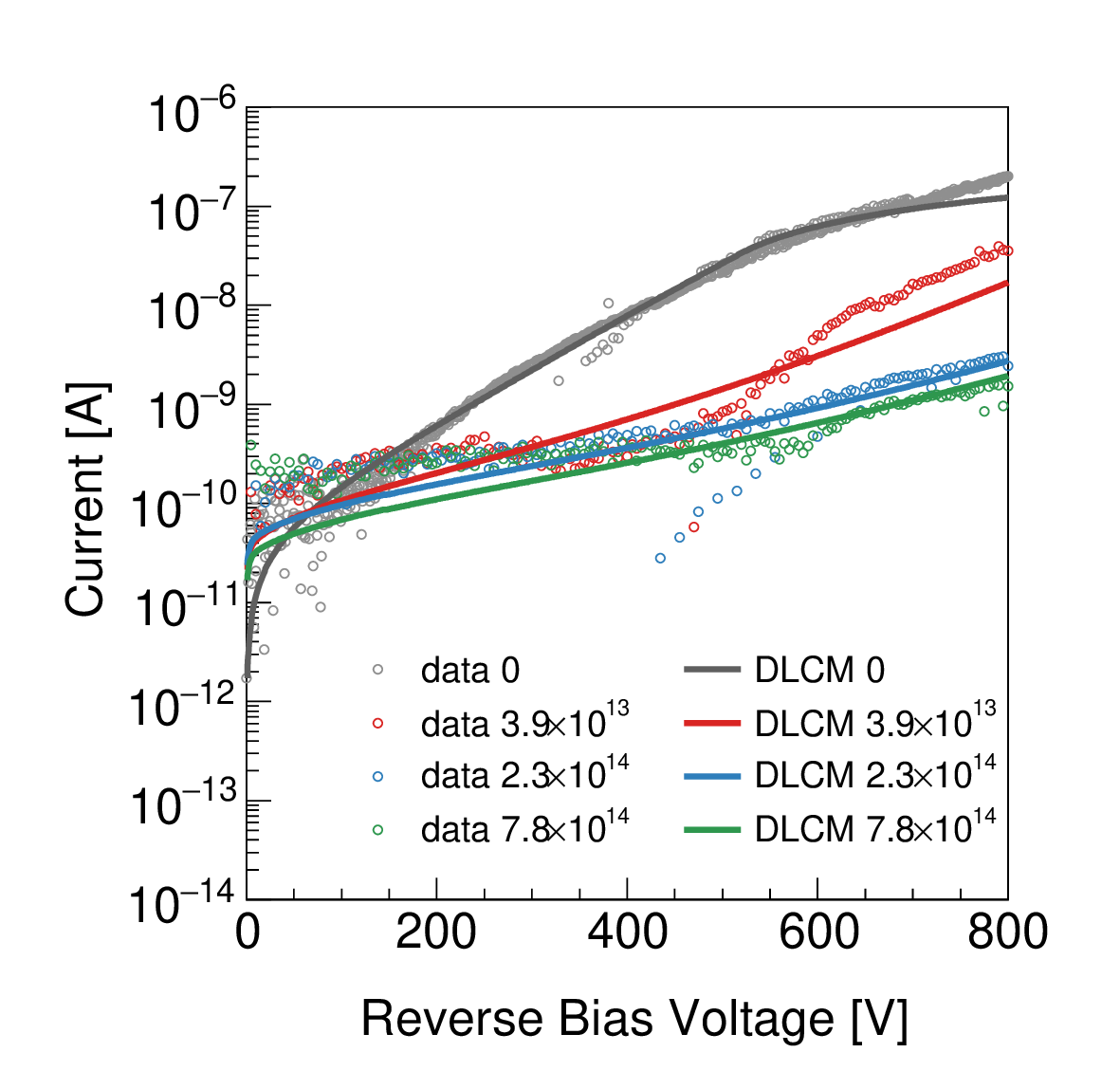}
    \caption{Measured and simulated I-V by DLCM for diodes irradiated from 0 to $7.8\times10^{14} n_{eq}/cm^2$ }
    \label{fig:iv-sim}
\end{figure}
\par Fig. \ref{fig:iv-sim} shows the measured current-voltage characteristics and simulation results. For higher voltages the simulations reproduce the I-V curves approximately over the fluence range whereas for the lower voltage the leakage current is below the experimental limit of resolution. The agreement of the results indicates that the assumptions made in DLCM are valid, and the reduction in leakage current after irradiation is primarily affected by the change in electric field, as illustrated in $\eqref{eq:G-R def}$.

\section{Conclusion}
In this study, irradiated $1.5\ mm\times1.5\ mm$ and $5\ mm\times5\ mm$ PINs are tested to assess their degradation after irradiation. By combining measured I-V and C-V results with established theories on radiation effects, we 
studied two main physical 
mechanism compensation effect and recombination process that significantly impact electrical properties. 
The DLCM simulation model was constructed using input parameters derived from DLTS and TRPL test results. To describe the change in doping concentration after irradiation, we introduced the g factor and developed an equivalent physical structure for C-V characteristic simulation, the results of which show good agreement with experimental data.
The g-factor will be further optimized with additional experimental data in subsequent studies.
The tunneling process, along with modifications to the carrier lifetime and effective intrinsic carrier concentration, is used in the simulation of the I-V characteristics, which demonstrate good agreement with the experimental data. Therefore, a new simulation model has been developed to predict the electrical properties of 4H-SiC devices for high-energy physics applications, demonstrating excellent agreement with experimental data. By constructing a DLCM physical model, the post-irradiation electrical performance of 4H-SiC devices can be compared within a unified framework to analyze irradiation-induced degradation mechanisms and evaluate radiation hardness, which is crucial for designing radiation-hardened devices. 

\newpage
\section{Appendix \MakeUppercase{\expandafter{\romannumeral1}}: equations for TCAD simulation}
TCAD device simulation employs a semi-classical approach to electronic transport in semiconductors. The set of equations utilized includes the Poisson equation $\eqref{eq-poisson}$, which describes the electric field in relation to the instantaneous charge density, along with the current continuity equations in a semiconductor $\eqref{eq-n-current}$, $\eqref{eq-p-current}$, where we take into account the contribution of carrier drift and diffusion. The simulation process solves the Poisson equation and current continuity equations \cite{M3}:
\begin{equation}
    \nabla\cdot\Vec{E}=\frac{\rho}{\epsilon_s}\label{eq-poisson}
\end{equation}
\begin{equation}
    \frac{\partial n}{\partial t}=G_n-R_n+\frac{1}{q}\nabla\cdot\Vec{J_n}\label{eq-n-current}
\end{equation}
\begin{equation}
    \frac{\partial p}{\partial t}=G_p-R_p+\frac{1}{q}\nabla\cdot\Vec{J_p}\label{eq-p-current}
\end{equation}
Considering the drift component proportional to electric field and the diffusion component proportional to the gradient of carrier concentration, the form of drift-diffusion currents in $\eqref{eq-n-current}$, $\eqref{eq-p-current}$ can be written as:
\begin{equation}
    \Vec{J_{n}}=qn\mu_n\Vec{E}+qD_n\nabla n\label{eq-jn}
\end{equation}
\begin{equation}
    \Vec{J_{p}}=qp\mu_p\Vec{E}+qD_p\nabla p\label{eq-jp}
\end{equation}
\section{Appendix \MakeUppercase{\expandafter{\romannumeral2}}: Supporting Figures}
\begin{figure}[H]
    \centering
    \includegraphics[width=1.\linewidth]{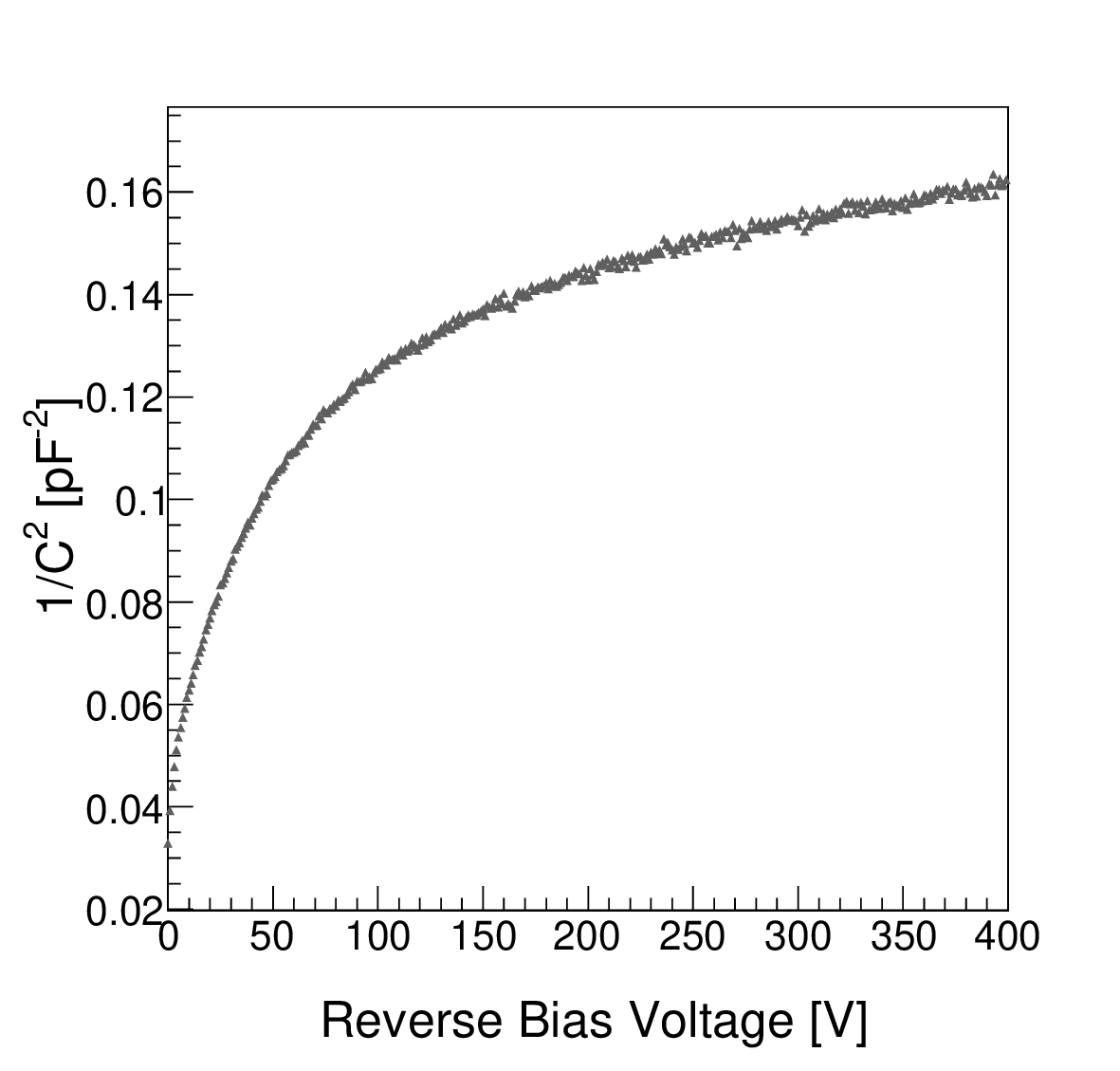}
    \caption{$1/C^2-V$ curve of non-irradiated $1.5\ mm\times1.5\ mm$ diode.}
    \label{fig:enter-label}
\end{figure}
\begin{figure}[H]
    \centering
    \includegraphics[width=1.\linewidth]{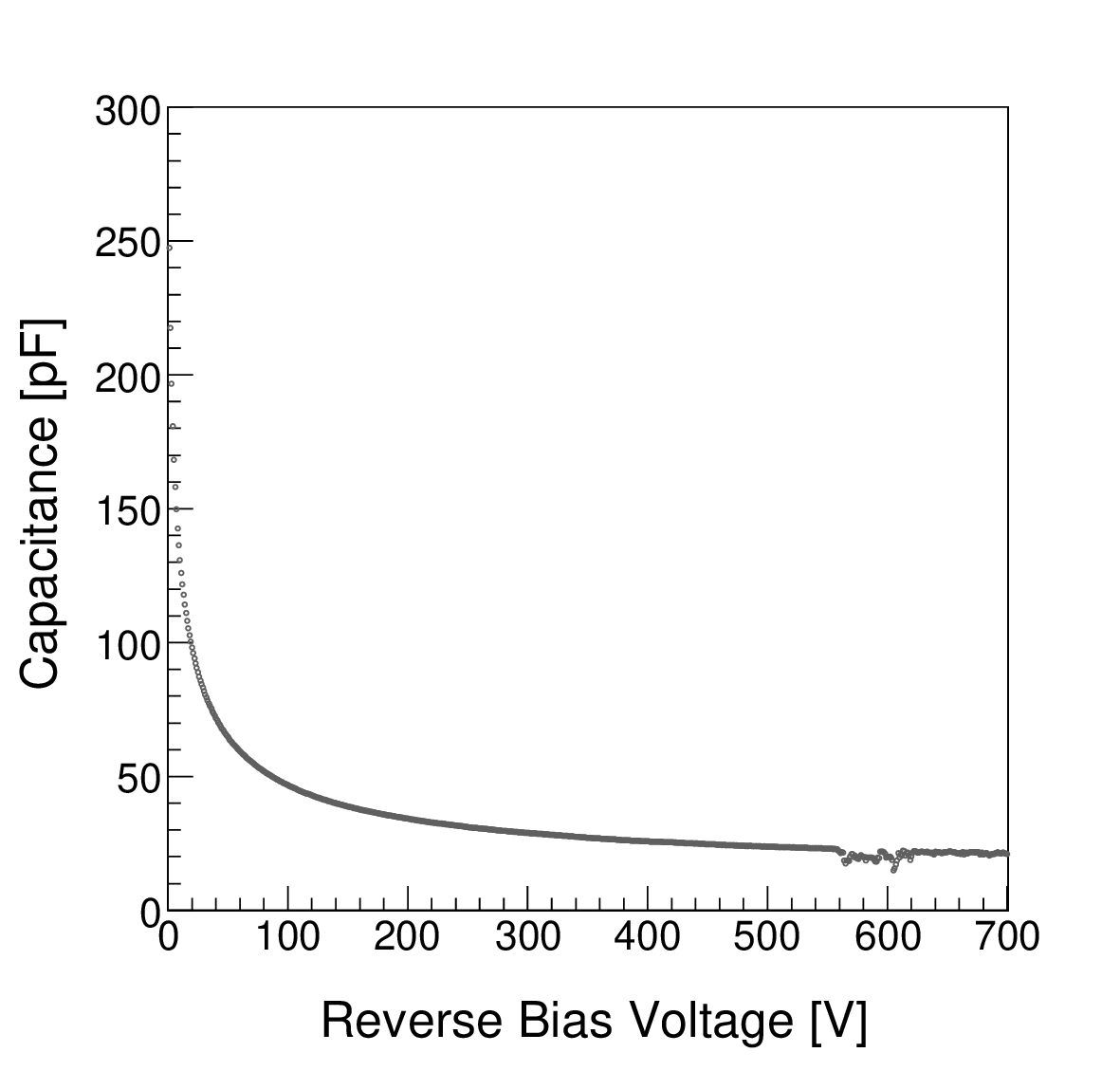}
    \caption{Total C-V characteristic of non-irradiated $5\ mm\times5\ mm$ diode.}
    \label{fig:enter-label}
\end{figure}
\begin{figure}[H]
    \centering
    \includegraphics[width=1.\linewidth]{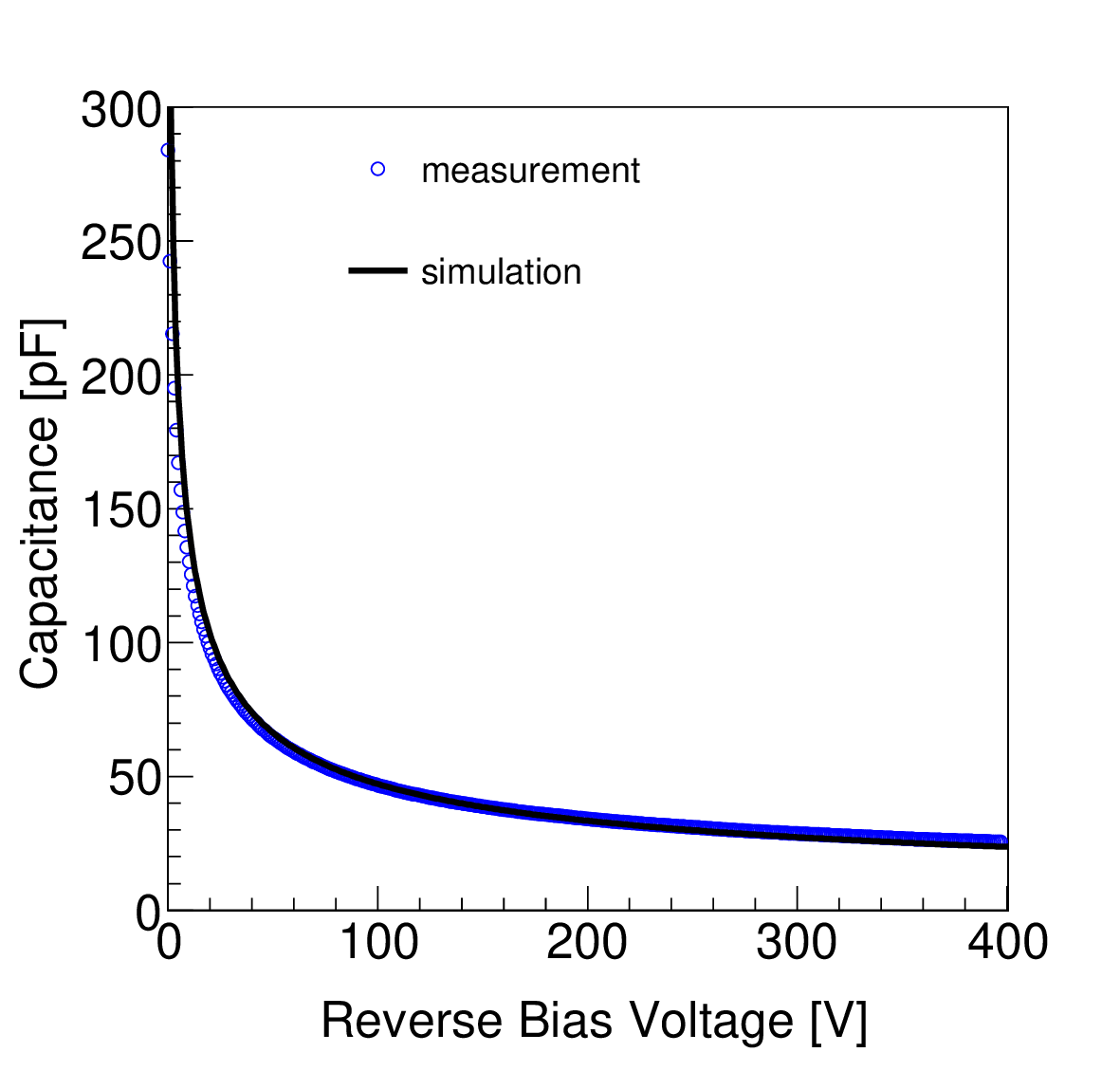}
    \caption{Simulated C-V of non-irradiated $5\ mm\times5\ mm$ diode.}
    \label{fig:enter-label}
\end{figure}
\end{document}